\documentclass[3p,times,procedia]{elsarticle}
\usepackage{nupha_ecrc}
\usepackage{graphicx}
\usepackage{subcaption}
\usepackage{amsmath}
\usepackage{slashed}

\newcommand{\beq}{\begin{equation}}
\newcommand{\eeq}{\end{equation}}
\newcommand{\ret}{\mathrm{ret}}
\newcommand{\adv}{\mathrm{adv}}

\volume{00}

\firstpage{1}

\journalname{Nuclear Physics A}

\runauth{S. Hauksson et al.}

\jid{nupha}

\jnltitlelogo{Nuclear Physics A}

\usepackage{amssymb}

\usepackage[figuresright]{rotating}

\begin{document}

\begin{frontmatter}



\dochead{XXVIIth International Conference on Ultrarelativistic Nucleus-Nucleus Collisions\\ (Quark Matter 2018)}

\title{Penetrating probes: Jets and photons in a non-equilibrium quark-gluon
plasma}


\author{Sigtryggur Hauksson, Sangyong Jeon, Charles Gale}

\address{Department of Physics, McGill University, 3600 University Street, Montreal, QC, H3A 2T8, Canada}

\begin{abstract}
We employ new field-theoretical tools to study photons and jets in a non-equilibrium quark-gluon plasma. Jet broadening and photon emission takes place through radiation which is suppressed by repeated and coherent interaction with the medium. We analyze this physics in an anisotropic plasma such as is created in the early stages of heavy-ion collisions. The anisotropy introduces an angular dependence in radiation and reduces its overall rate. This can affect phenomenological predictions of the rapidity dependence and angular flow of jets and photons.  
\end{abstract}


\end{frontmatter}


\section{Introduction}
\label{}


Heavy-ion collisions allow us to study the non-equilibrium physics of quark-gluon plasma (QGP). 
They give us access to transport coefficients of the QGP, such as its shear viscosity, and tell us how the far-from-equilibrium medium created in the early stages of collisions evolves towards a hydrodynamic description. To understand these non-equilibrium phenomena, sensitive experimental probes are needed as well as theoretical calculations of how deviations from thermal equilibrium affect the probes. Jets and photons are especially suitable because they are responsive to the whole evolution of the medium.

\begin{figure}
    \centering
    \begin{subfigure}[b]{0.2\textwidth}
        \includegraphics[width=\textwidth]{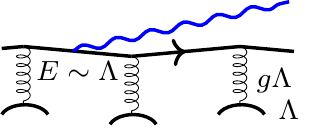}
        \caption{}
        \label{Fig:LPM_photon}
    \end{subfigure}
    \quad\quad
    \begin{subfigure}[b]{0.2\textwidth}
        \includegraphics[width=\textwidth]{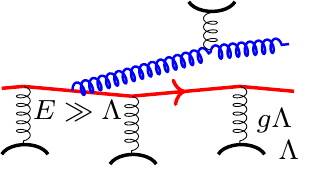}
        \caption{}
        \label{Fig:LPM_jet}
    \end{subfigure}
    \quad\quad
    \begin{subfigure}[b]{0.2\textwidth}
        \includegraphics[width=\textwidth]{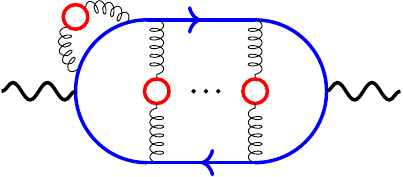}
        \caption{}
        \label{Fig:cut}
    \end{subfigure}
    \caption{Diagrams for medium-induced bremsstrahlung. The hard scale is \(\Lambda\) and the soft scale is \(g\Lambda\) with \(g\) the QCD coupling constant. Figure (a) shows a photon radiating off a quark in the medium; (b) shows a gluon radiating off a quark in a jet; (c) is an example of the diagrams needed to evalute photon emission. It can be shown to correspond to the amplitude in (a) squared by using cutting rules. In all cases an arbitrary number of interactions with the medium can take place during the radiation.}
\end{figure}

The physics of jets and photons in a weakly coupled plasma shares many similarities. Photons are radiated in two-to-two scattering and in medium-induced bremsstrahlung and quark-antiquark annihilation as can be seen in Fig. \ref{Fig:LPM_photon}. Similarly, jet particles can interact with the medium in two-to-two scatterings and radiate  a gluon in medium-induced bremsstrahlung, see Fig. \ref{Fig:LPM_jet}. During both photon and gluon radiation through bremsstrahlung the radiating particle  interacts repeatedly and coherently with medium. This is known as the Landau-Pomeranchuk-Migdal (LPM) effect and tends to reduce the emission rate \cite{AMYphotonnumerics}. 

In these proceedings we present first results on medium-induced bremsstrahlung in a non-equilibrium medium. For jets this channel dominates the physics and for photon production it is as important as two-to-two scattering which has been evaluated out of equilibrium in e.g. \cite{Schenke2006,Hauksson2016}. We derive general equations describing the bremsstrahlung, taking the LPM effect fully into account. We furthermore solve these equations in the special case of an anisotropic medium characterized by a momentum distribution of the form \cite{Romatschke2003} 
\beq
\label{Eq:aniso}
f(\mathbf{p}) = \sqrt{1+\xi}\; f_{\mathrm{eq}}\left(\sqrt{p^2 + \xi (\mathbf{p} \cdot \mathbf{n})^2}/\Lambda \right).
\eeq
Here \(f_{\mathrm{eq}}\) is an equilibrium distribution (either Fermi-Dirac or Bose-Einstein) which is deformed by an anisotropy \(\xi\) in the direction of \(\mathbf{n}\). The hard scale \(\Lambda\) corresponds to the temperature in equilibrium. The factor \(\sqrt{1+\xi}\) enforces a constant number of particles in and out of equilibrium. 

\section{Theoretical development}

We use tools of out-of-equilibrium field theory to derive equations describing photon production through medium-induced bremsstrahlung \cite{Hauksson2017}. 
A detailed derivation of jet splitting will be reported elsewhere \cite{inprep}. 
Using the real-time formalism \cite{Calzetta1986} we define
\(
\psi_r = \frac{1}{2} \left( \psi_1 + \psi_2 \right)\)
and
\(
 \psi_a = \psi_1 - \psi_2
\)
and consider the correlators of these fields.
The fundamental assumption of our calculation is that the momentum distribution varies slowly in space and time. For quarks this allows us to write 
\beq
\label{Eq:barerr}
S^0_{rr}(P)  = \left[ \frac{1}{2} - F(P)\right] \left(S^0_{ra} - S^0_{ar} \right).
\eeq
where all correlators are bare and 
\beq
F(P) =
\begin{cases}
f(\mathbf{p}), &  p^0>0 \\
 1-f(\mathbf{p}), &  p^0 < 0
\end{cases}
\eeq
describes the momentum distribution \(f\) including Pauli blocking \cite{Calzetta1986}. 

The two momentum scales of our problem are those of soft gluon mediators with momentum \(\mathcal{O}(g\Lambda)\) and those of hard quasiparticles which have momentum \(P \sim \Lambda\) and are almost on shell, \(P^2 \sim g^2 \Lambda^2\). The retarded, resummed correlator of the hard quasiparticles is 
\beq
S_{\ret} = \frac{i \slashed{P}}{P^2 - m^2 - i \Gamma p^0}
\eeq 
which describes quasiparticles with thermal mass \(m\) and decay width \(\Gamma\). Their \(rr\) correlator is \cite{Hauksson2017}  
\beq
S_{rr}(P)  = \left[ \frac{1}{2} - F(P)\right] \left(S_{\ret} - S_{\adv} \right).
\eeq
This is identical to Eq. \eqref{Eq:barerr}, except all correlators are now resummed. Thus the hard quasiparticles have the same momentum distribution as the bare particles. For the soft, mediating gluons we have provided a first calculation of their density in a non-equilbrium medium by evaluting their \(G_{rr}\) correlator using the anisotropic momentum distribution in Eq. \eqref{Eq:aniso} \cite{inprep}.

An example of the diagrams that need to be evaluted is shown in Fig. \ref{Fig:cut}. There are infinitely many such diagrams with many ways of placing \(r\) and \(a\) indices. Remarkably, a careful examination \cite{Hauksson2017} shows that almost all of these diagrams cancel or are subleading allowing for a simple resummation of the remaining diagrams. This derivation is general since it does not invoke the Kubo-Martin-Schwinger relation which relates n-point functions in thermal equilibrium. The final result is consistent with the results from kinetic theory in \cite{AMYkinetic}. The differential rate for photon production, \(R\), is given by
\beq
\label{Eq:rate}
k \frac{dR}{d^3k} = \frac{3Q^2 \alpha_{EM}}{4 \pi^2} \int \frac{d^3 p}{(2\pi)^3} F(P+K) \left[ 1- F(P)\right] \frac{p^{z\;2} + (p^z + k)^2}{2 p^{z\;2} (p^z + k)^2} \mathbf{p_{\perp}} \cdot \mathrm{Re} \;\mathbf{g}(\mathbf{p_{\perp}})
\eeq 
where \(\mathbf{p_{\perp}}\) lies in the plane perpendicular to the direction of the photon momentum, \(\mathbf{k} = k \mathbf{e}_z\). The function \(\mathbf{g}(\mathbf{p_{\perp}})\) is a solution of the integral equation
\beq
\label{Eq:integral_eq}
 \mathbf{p_{\perp}} = i \delta E\; \mathbf{g}(\mathbf{p_{\perp}}) + \int \frac{d^2 q_{\perp}}{(2\pi)^2} \; \mathcal{C}(\mathbf{q}_{\perp}) \left[\mathbf{g}(\mathbf{p_{\perp}}) - \mathbf{g}(\mathbf{p_{\perp}} + \mathbf{q}_{\perp})\right].
\eeq
The collision kernel, \(\mathcal{C}\), describes the density of mediating gluons that the emitting quark sees. It can be obtained from \(G_{rr}\) \cite{Hauksson2017}.

The assumption that the momentum distribution in Eq. \eqref{Eq:barerr} varies slowly in space and time can break down in an anisotropic medium. Specifically, Weibel instabilities give rise to strong and rapidly growing chromomagnetic fields \cite{Mrowczynski2016}. This instability corresponds to a pole in the retarded gluon correlator, \(G_{ra}(p^0)\), in the upper half of the complex plane \cite{Romatschke2003}. Because of this pole the collision kernel, \(\mathcal{C}\),  diverges. To deal with this problem we assume that the anisotropy is small, \(\xi < g^2\), and cut off the low-energy part of the collision kernel in which the divergence occurs.  We then expand seperately in \(g\) and \(\xi\).  This is justified because for small anisotropy the chromomagnetic fields never become very strong and their growth is slower than \(1/g^2 \Lambda\) \cite{Kurkela2011}, the time scale of the LPM effect. Work on a theoretically consistent solution for larger anisotropies is ongoing.

\section{Results and implications for phenomenology}

\begin{figure}
    \centering
    \begin{subfigure}[b]{0.32\textwidth}
        \includegraphics[width=\textwidth]{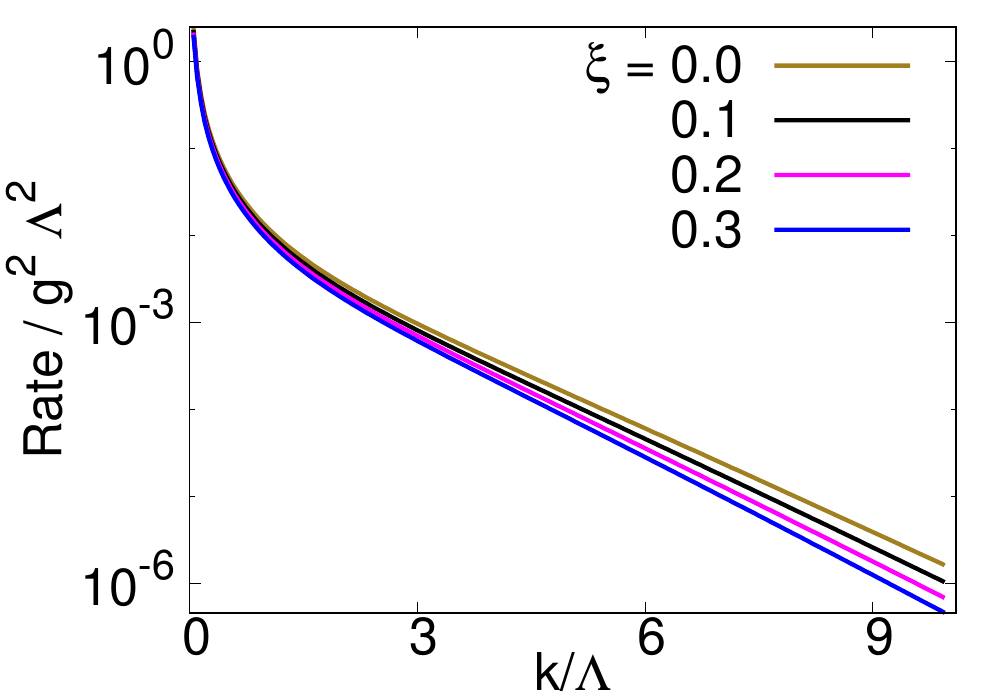}
        \caption{Rate for \(\theta=0\).}
        \label{Fig:photon_rate}
    \end{subfigure}
    \begin{subfigure}[b]{0.32\textwidth}
        \includegraphics[width=\textwidth]{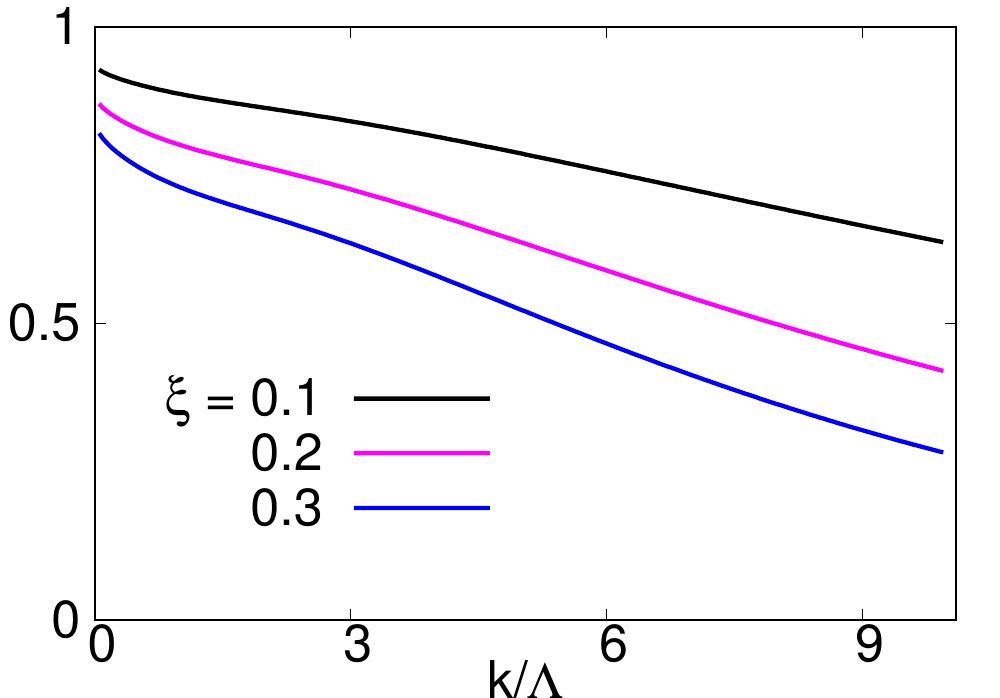}
        \caption{Ratio with \(\xi=0\) at \(\theta=0\).}
        \label{Fig:photon_ratio}
    \end{subfigure}
    \begin{subfigure}[b]{0.32\textwidth}
    	\includegraphics[width=\textwidth]{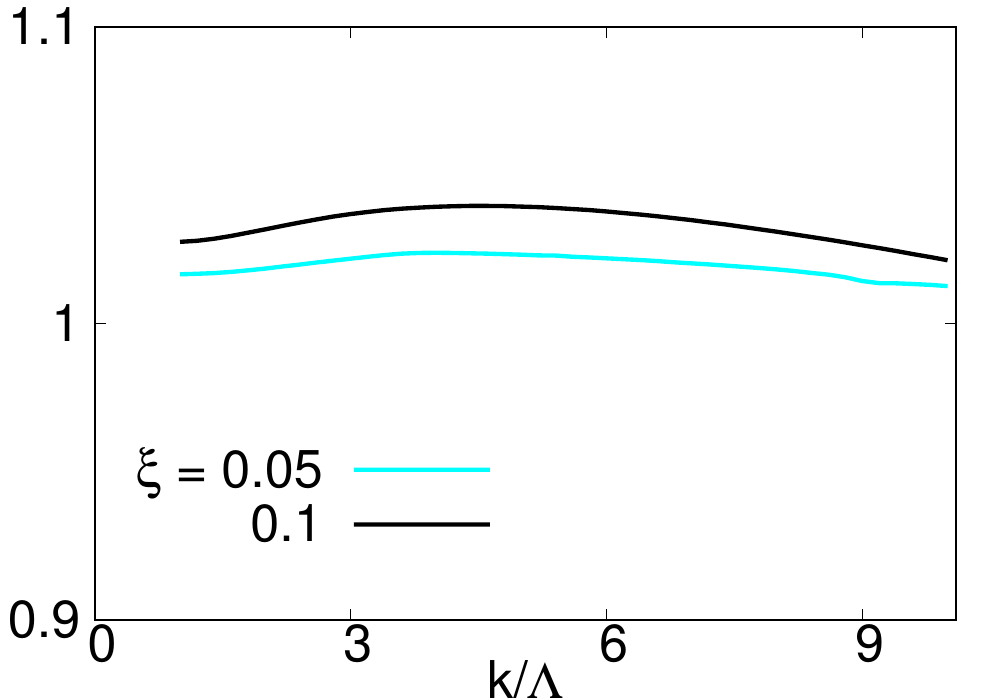}
        \caption{Ratio with \(\xi=0\) at \(\theta=1.2\).}
        \label{Fig:photon_theta1p2}
    \end{subfigure}
    \caption{Rate of photon emission through medium-induced bremsstrahlung in an anisotropic medium. The rate is reduced for emission in the direction of the anisotropy, \(\theta = 0\), and slightly increased for emission nearly orthogonal to the anisotropy \(\theta = 1.2\). These effects are more pronounced at higher anisotropy \(\xi\).}
\end{figure}

We now evaluate the rate of photon production through medium-induced bremsstrahlung in an anisotropic medium as in Eq. \eqref{Eq:aniso}. This is obtained by solving Eqs. \eqref{Eq:rate} and \eqref{Eq:integral_eq} numerically by expanding \(\mathbf{g}\) in a functional basis.  The rate depends on the strength of the anisotropy \(\xi\) and the angle \(\theta\) between the vector \(\mathbf{n}\) and the momentum of the photon.
 Figs. \ref{Fig:photon_rate} and \ref{Fig:photon_ratio} show the rate for different values of \(\xi\) at \(\theta=0\). 
 We see that the rate is reduced significantly as the anisotropy increases. Fig. \ref{Fig:photon_theta1p2} shows the rate when the photon momentum and \(\mathbf{n}\) are nearly orthogonal to each other, \(\theta = 1.2\). The rate is slightly increased with increasing anisotropy.  

These results come about through two effects: Firstly the collision kernel changes in a non-equilibrium medium. In thermal equilibrium it is \cite{Aurenche2002}
\beq
\mathcal{C}(\mathbf{q}) = C_F \left[\frac{1}{q^2} - \frac{1}{q^2 + m_D^2} \right].
\eeq 
where \(m_D^2\) is the Debye mass.
This expression diverges when \(\mathbf{q} \rightarrow 0\) showing that the quark emitter and the medium mostly interact through very low energy gluons.
 In an anisotropic medium this low-energy divergence is screened. Schematically, our calculation shows that
\(
1/q^2 \rightarrow 1/(q^2 + \xi m^2)
\)
where \(q^2\) and \(m^2\) are of order \(g^2 \Lambda^2\). This means that the total rate of photon emission is reduced because there are fewer interactions with the medium to stimulate emission.
The second important effect is the angular distribution of the emitting quarks. Simply put, there are fewer collinearly emitted photons when there are fewer quarks travelling in a certain direction. 
These two effects both reduce emission for \(\theta \sim 0\) while they compete for \(\theta \sim \pi/2\) leading to a slightly enhanced emission rate. 

These results could have important implications for phenomenology. The QGP medium formed in heavy-ion collisions is initally highly anisotropic in the rapidity direction. Thus, there should be fewer photons emitted closer to the beamline, reducing the number of photons at higher rapidity. Our calculation could also play a role in the elliptic flow of photons since viscous hydrodynamics assumes non-isotropic momentum distributions which vary within an event and between events.  Understanding the importance of these effects requires further studies using realistic hydrodynamical simulations.

\begin{figure}
    \centering
    \begin{subfigure}[b]{0.32\textwidth}
        \includegraphics[width=\textwidth]{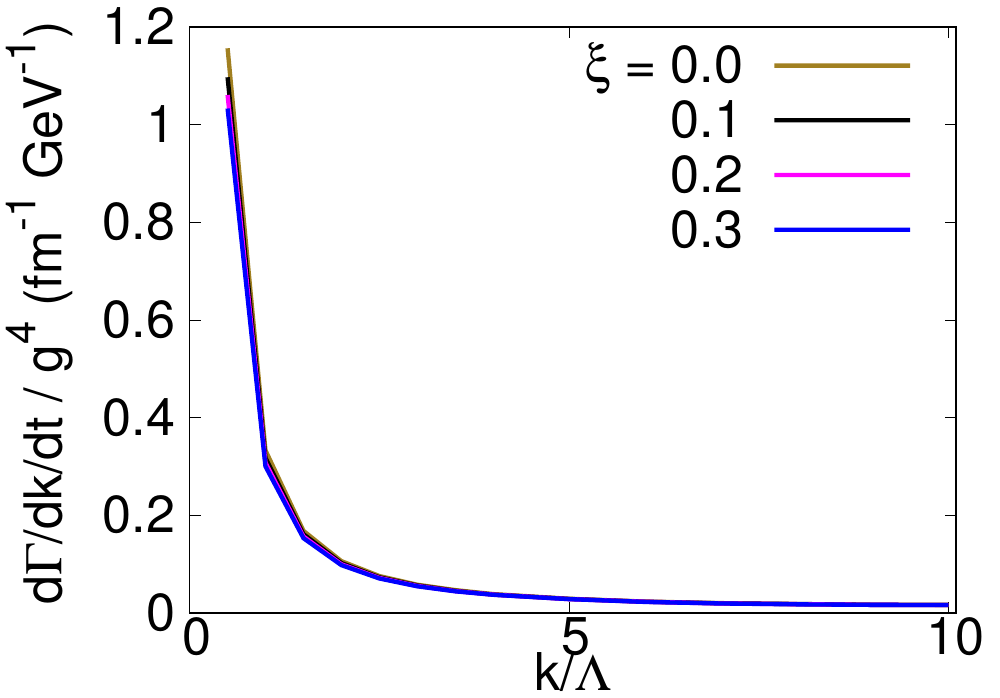}
        \caption{Radiation rate.}
        \label{Fig:jet_rate}
    \end{subfigure}
    \quad\quad
    \begin{subfigure}[b]{0.32\textwidth}
        \includegraphics[width=\textwidth]{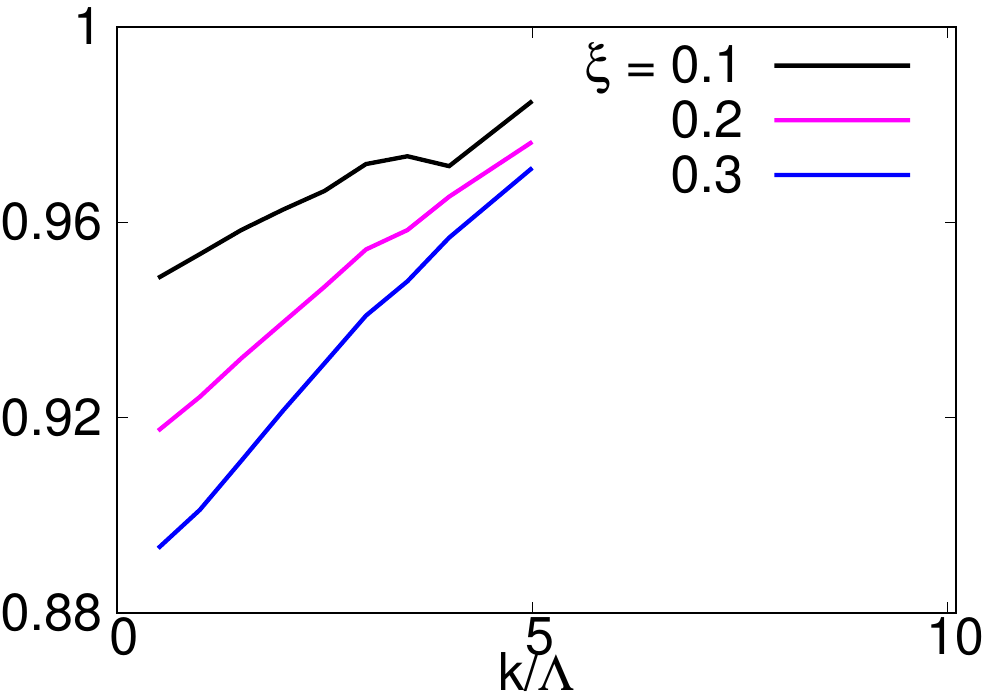}
        \caption{Ratio with \(\xi=0\).}
        \label{Fig:jet_ratio}
    \end{subfigure}
    \caption{Rate of gluon radiation off a jet gluon with energy \(20\Lambda\) through medium-induced bremsstrahlung. We assume an anisotropic medium and consider emission in the direction of the anisotropy, \(\theta = 0\). The rate is reduced, especially when the radiated gluon has low energy.}\label{Fig:jet_calc}
\end{figure}

We have also evaluated the rate for a jet particle to radiate a gluon through medium-induced bremsstrahlung. Fig. \ref{Fig:jet_calc} shows this rate for a gluon of energy \(20\Lambda\) which radiates a gluon of energy \(k\). This is shown for the case of \(\theta=0\), i.e. all particles moving in the direction of \(\mathbf{n}\). As before, the rate is reduced at higher anisotropy because interaction through very low energy mediators is screened. This effect is strongest when the emitted gluon has fairly low momentum. These low momentum gluons are absorbed by the medium suggesting that at higher anisotropy jet suppression is reduced in the direction of the anisotropy. This could affect the rapidity dependence of jet particles and their directionality in the transverse plane.

%

 \appendix
 
 \vfill
 {\noindent }\textbf{Acknowledgment:} This work was supported in part by the Natural Sciences and Engineering Research Council of Canada. CG gratefully acknowledges the Canada Council for the Arts for funding through its Killam Research Fellowship Program. SH acknowledges support through grants from Fonds de Recherche du Qu\'ebec.



\bibliographystyle{elsarticle-num}
\bibliography{bibliography.bib}







\end{document}